\documentclass[12pt]{article}
\usepackage{epsf}
\title{ {\bf
Lepton flavor conserving $Z\rightarrow l^+ l^-$ decays in the
general two Higgs doublet model}}
\author{\vspace{1cm}\\
        {\bf E. O. Iltan}
        \thanks{E-mail address:
        eiltan@heraklit.physics.metu.edu.tr}
\\
        Physics Department, Middle East Technical University \\
        Ankara, Turkey\\}

\date{}

\begin{document}
\setlength{\baselineskip}{24pt}
\maketitle
\setlength{\baselineskip}{7mm}
\begin{abstract}
We calculate the new physics effects to the branching ratios of the lepton
flavor conserving decays $Z\rightarrow l^+ l^-$ in the framework of the
general two Higgs Doublet model. We predict the upper limits for the
couplings $|\bar{\xi}^{D}_{N,\mu\tau}|$ and $|\bar{\xi}^{D}_{N,\tau\tau}|$
as $3\times 10^2\,GeV$ and $1\times 10^2\,GeV$, respectively.  
\end{abstract} 
\thispagestyle{empty}
\newpage
\setcounter{page}{1}
\section{Introduction}
In the standard model (SM) of electroweak interactions lepton flavor is
conserved. This conservation can be broken with the extension of the SM, 
such as $\nu$SM, permitting the existence of the massive neutrinos and 
the lepton mixing mechanism \cite{Pontecorvo}, $\nu$SM, extended with one 
heavy ordinary Dirac neutrino or two heavy right-handed singlet Majorana 
neutrinos \cite{Illana}, Zee model \cite{Ghosal}, the general two Higgs 
doublet model, which contains off diagonal Yukawa couplings in the lepton 
sector \cite{iltanZl1l2}.  

Leptonic Z-decays are among the most interesting lepton flavor conserving 
(LFC) and lepton flavor violating (LFV) interactions and  they reached great 
interest since the related experimental measurements are improved at present.  
With the Giga-Z option of the Tesla project, there is a possibilty to
increase Z bosons at resonance \cite{Hawkings}.

The processes $Z\rightarrow l^- l^+$ with $l=e,\mu,\tau$ are the among the 
LFC decays and they exist in the SM even at the tree level. The experimental
predictions for the branching ratios of these decays are \cite{PartData}   
\begin{eqnarray}
BR(Z\rightarrow e^- e^+) &=& 3.366 \pm 0.0081\,\%  
\nonumber \, , \\
BR(Z\rightarrow \mu^- \mu^+) &=& 3.367 \pm 0.013\,\% 
\nonumber \, , \\ 
BR(Z\rightarrow \tau^- \tau^+) &=& 3.360 \pm 0.015 \,\% \, ,
\label{Expr1}
\end{eqnarray}
and the tree level SM predictions are  
\begin{eqnarray}
BR(Z\rightarrow e^- e^+) &=& 3.331\,\%  
\nonumber \, , \\
BR(Z\rightarrow \mu^- \mu^+) &=& 3.331\,\% 
\nonumber \, , \\ 
BR(Z\rightarrow \tau^- \tau^+) &=& 3.328 \,\% \, .
\label{Expr2}
\end{eqnarray}
Comparision of these experimental and theoretical results shows that 
the main contribution comes from the SM in the tree level and the loop
contributions, even the ones beyond the SM, should lie almost in the 
uncertainity of the measurements of these decays.  
 
In the literature, there are various experimental and theoretical studies 
\cite{Kamon}-\cite{Achim}. A method to determine the weak electric dipole 
moment was developed in \cite{Botz}. The vector and axial coupling 
constants, $v_f$ and $a_f$, in Z-decays have been measured at LEP 
\cite{LEP}. Furthermore, the measurements of the weak electric dipole 
moments of fermions have been performed  \cite{WEDM}. In \cite{Stiegler}, 
various additional types of interactions have been studied and a way to 
measure these contributions in the process $Z\rightarrow \tau^- \tau^+$ 
was described. In \cite{Achim}, a new method to measure the electroweak 
mixing angle in Z-decays to tau leptons has been proposed.

In the present work we study the LFC $Z\rightarrow l^- l^+$ decays, 
where $ l=e,\mu,\tau$, in the model III version of 2HDM, which is the
minimal extension of the SM. Since the SM prediction for these decays, 
even in the tree level, is almost the same as the experimental one, 
the contributions beyond the SM, which can exist at least in the loop
level, should be small enough not to exceed the experimental results. 
This discussion stimulates us to study the BR's of these processes with
the addition of the contributions beyond the SM and try to predict upper
limits for the new couplings existing in the model used. In the model 
III, the neutral Higgs bosons $h^0$ and $A^0$ play the important role
for the physics beyond the SM, in the calculation of the BR of the LFC
decays underconsideration. This analysis shows that the predictions of  
the upper limits for the couplings $|\bar{\xi}^{D}_{N,\mu\tau}|$ and
$|\bar{\xi}^{D}_{N,\tau\tau}|$ are $\sim 3\times 10^2\,GeV$ and
$\sim 1\times 10^2\, GeV$, respectively, however, an upper limit for the 
coupling $|\bar{\xi}^{D}_{N,e\tau}|$ can not be found due to the small
contribution of new physics effects to the BR of the $Z\rightarrow e^-
e^+ $ decay. 
 
The paper is organized as follows:
In Section 2, we present the explicit expressions for the branching ratios
of $Z\rightarrow l^- l^+$ in the framework of the model III. Section 3 is 
devoted to discussion and our conclusions.
\section{$Z\rightarrow l^- l^+$ decay in the general two Higgs Doublet 
model.} 
In the SM, lepton flavor is conserved since the matter content forbids the
lepton flavor violation. However, most theories beyond the SM may bring
flavor changing neutral currents (FCNC) at the tree level, unless some 
discrete ad hoc symmetries are imposed to eliminate them.  The model I 
and II versions of 2HDM are the examples of the theories beyond where FCNC 
at the tree level is forbidden. In the model III version of 2HDM, the 
FCNC interactions at the tree level are allowed and this makes the LFV 
interactions possible. Furthermore, the existence of FCNC at the tree level
brings new contributions to LFC decays. The most general Yukawa interaction 
for the leptonic sector in the model III is
\begin{eqnarray}
{\cal{L}}_{Y}=
\eta^{D}_{ij} \bar{l}_{i L} \phi_{1} E_{j R}+
\xi^{D}_{ij} \bar{l}_{i L} \phi_{2} E_{j R} + h.c. \,\,\, ,
\label{lagrangian}                                                                                            
\end{eqnarray}
where $i,j$ are family indices of leptons, $L$ and $R$ denote chiral 
projections $L(R)=1/2(1\mp \gamma_5)$, $\phi_{i}$ for $i=1,2$, are the 
two scalar doublets, $l_{i L}$ and $E_{j R}$ are lepton doublets and
singlets respectively. The choice of $\phi_{1}$ and $\phi_{2}$ 
\begin{eqnarray}
\phi_{1}=\frac{1}{\sqrt{2}}\left[\left(\begin{array}{c c} 
0\\v+H^{0}\end{array}\right)\; + \left(\begin{array}{c c} 
\sqrt{2} \chi^{+}\\ i \chi^{0}\end{array}\right) \right]\, ; 
\phi_{2}=\frac{1}{\sqrt{2}}\left(\begin{array}{c c} 
\sqrt{2} H^{+}\\ H_1+i H_2 \end{array}\right) \,\, ,
\label{choice}
\end{eqnarray}
with the vacuum expectation values   
\begin{eqnarray}
<\phi_{1}>=\frac{1}{\sqrt{2}}\left(\begin{array}{c c} 
0\\v\end{array}\right) \,  \, ; 
<\phi_{2}>=0 \,\, ,
\label{choice2}
\end{eqnarray}
helps us to decompose the SM particles and beyond in the tree level. In this
case, the first doublet carries the SM particles and the other one is
responsible for the particles beyond the SM. Therefore, we take $H_1$ 
and $H_{2}$ (see eq. (\ref{choice})) as the mass eigenstates $h^0$ and $A^0$ 
respectively, since no mixing between CP-even neutral Higgs bosons $h^0$ and 
the SM one, $H^0$, occurs at the tree level. In eq. (\ref{lagrangian}) 
$\xi^{D}_{ij}$  are the Yukawa matrices and they have in general complex 
entries. Notice that in the following we replace $\xi^{D}$ with $\xi^{D}_{N}$ 
where "N" denotes the word "neutral". 

The general effective vertex for the interaction of on-shell Z-boson with a
fermionic current is given by
\begin{eqnarray}
\Gamma_{\mu}=\gamma_{\mu}(f_V-f_A\ \gamma_5)+ \frac{1}{m_W} (f_M+f_E\,
\gamma_5)\, \sigma_{\mu\,\nu}\, q^{\nu}
\label{vertex}
\end{eqnarray}
where $q$ is the momentum transfer, $q^2=(p-p')^2$, $f_V$ ($f_A$) is vector 
(axial-vector) coupling, $f_M$ ($f_E$) is proportional to the weak magnetic 
(electric dipole) moment of the fermion. Here $p$ ($-p^{\prime}$) is the four 
momentum vector of lepton (anti-lepton). For the $Z\rightarrow l^- l^+$ decay, 
the couplings $f_V$ and  $f_A$ have contributions from the SM, $f^{SM}_V$ and 
$f^{SM}_A$, even at the tree level, and all the couplings have contributions 
beyond the SM, $f^{Beyond}_I$, where $I=V,A,M,E$. The explicit expressions 
for these couplings are 
\begin{eqnarray}
f^{SM}_V&=& \frac{i g }{cos\,\theta_W}\, c_V \, \nonumber \\
f^{SM}_A&=& \frac{i g }{cos\,\theta_W}\, c_A \, ,
\end{eqnarray}
and 
\begin{eqnarray}
f^{Beyond}_V&=& \frac{-i g}{32\,cos\,\theta_W\,\pi^2} 
\Bigg \{
\int_0^1\, dx \,  c_V\, \eta^V_i\, (-1+x)\, (ln \, \frac{L^{self}_{h^0}}
{\mu^2}\, \frac{L^{self}_{A^0}}{\mu^2} )
\nonumber \\ &+&
\int_0^1\,dx\, \int_0^{1-x} \, dy \, 
\Bigg (
\frac{1}{2}\,(-1+x+y)\,  m_i\, m_{l^-}\,\eta_i^- \, 
(\frac{1}{L^{ver}_{h^0\,A_0}}-\frac{1}{L^{ver}_{A^0\,h^0}}) 
\nonumber \\ &+& 
c_V  \Big [ 
(-\eta^V_i \, m_i^2 + (-1+x+y)\, \eta_i^+ \,m_i\, m_{l^-} )\, 
\frac{1}{L^{ver}_{h^0}} 
\nonumber \\ &-&
(\eta^V_i \, m_i^2 + (-1+x+y)\, \eta_i^+ \, m_i\, m_{l^-})\, 
\frac{1}{L^{ver}_{A^0}} 
\nonumber \\ &+& 
\eta^V_i \,(2-(q^2\,x\,y+m_{l^-}^2\, (-1+x+y)^2)\, (\frac{1}{L^{ver}_{h^0}} 
+ \frac{1}{L^{ver}_{A^0}})+ln\,\frac{L^{ver}_{h^0}}{\mu^2}\,
\frac{L^{ver}_{A^0}}{\mu^2} ) \Big ] \Bigg ) \Bigg \} \, , 
\nonumber \\
f^{Beyond}_A&=& \frac{i g}{32\,cos\,\theta_W\,\pi^2} 
\Bigg \{
\int_0^1\, dx \,  c_A\, \eta^V_i\, (-1+x)\, (ln \, \frac{L^{self}_{h^0}}
{\mu^2}\,\frac{L^{self}_{A^0}}{\mu^2} )
\nonumber \\ &-&
\int_0^1\,dx\, \int_0^{1-x} \, dy \, 
\Bigg (
-2\,c_A \,\eta_i^V \, ln\,\frac{L^{ver}_{h^0\, A^0}}{\mu^2}\,
\frac{L^{ver}_{A^0\,h^0}}{\mu^2} \nonumber \\ &+&
c_A  \Big [ 
(\eta^V_i \, m_i^2 - (-1+x+y)\, \eta_i^+ \,m_i\, m_{l^-} )\, 
\frac{1}{L^{ver}_{h^0}} 
\nonumber \\ &+&
(\eta^V_i \, m_i^2 + (-1+x+y)\, \eta_i^+ \, m_i\, m_{l^-})\, 
\frac{1}{L^{ver}_{A^0}} 
\nonumber \\ &+& 
\eta^V_i \,(2-(q^2\,x\,y-m_{l^-}^2\, (-1+x+y)^2)\, (\frac{1}{L^{ver}_{h^0}} 
+ \frac{1}{L^{ver}_{A^0}})+ln\,\frac{L^{ver}_{h^0}}{\mu^2}\,
\frac{L^{ver}_{A^0}}{\mu^2} ) \Big ] \Bigg ) \Bigg \} \, ,
\nonumber \\
f^{Beyond}_M&=& \frac{g\,m_W}{256\,cos\,\theta_W\,\pi^2} \,\,  
\int_0^1\,dx\, \int_0^{1-x} \, dy \, 
\Bigg \{
2\, \eta_i^- \, m_i\, (-1+x+y)\,(\frac{1}{L^{ver}_{h^0\,A^0}} - 
\frac{1}{L^{ver}_{A^0\,h^0}}) 
\nonumber \\ &+& 
\frac{1}{L^{ver}_{h^0}} \,  \Bigg (\eta_i^- \,m_i\,(y-x)-4\, c_V\, (x+y)\, 
\Big (-\eta_i^+ \, m_i + 2\,(-1+x+y)\,\eta_i^V\, m_{l^-}  \Big ) \Bigg ) 
\nonumber \\ &+&
\frac{1}{L^{ver}_{A^0}} \,  
\Bigg (\eta_i^- \,m_i\,(x-y)-4\, c_V\, (x+y)\, \Big ( \eta_i^+ \, m_i +
2\,(-1+x+y)\,\eta_i^V\, m_{l^-}  \Big ) \Bigg ) \Bigg \} \nonumber \, , \\
f^{Beyond}_E&=& \frac{g\,m_W}{256\,cos\,\theta_W\,\pi^2} \,\,  
\int_0^1\,dx\, \int_0^{1-x} \, dy \, 
\Bigg \{
2\,(1-x-y) \Bigg ( 
\nonumber \\
& & \Big ( \eta_i^+ \, m_i + 2\,(x-y)\,\eta_i^V\, m_{l^-} \Big )\, 
\frac{1}{L^{ver}_{h^0\,A^0}} - 
\Big ( \eta_i^+ \, m_i + 2\,(y-x)\,\eta_i^V\, m_{l^-} \Big )\, 
\frac{1}{L^{ver}_{A^0\,h^0}} 
\nonumber \\ &+& 
\Bigg ( m_i \Big ( \eta_i^+ \, (y-x)+4 \, c_V\, \eta_i^- \,(x+y) \Big ) 
+2\, (y-x)\, (-1+x+y)\,\eta_i^V \, m_{l^-} \Bigg ) \frac{1}{L^{ver}_{A^0}}
\nonumber \\ &+&
\Bigg ( m_i \Big ( \eta_i^+ \, (x-y)+4 \, c_V\, \eta_i^- \,(x+y) \Big ) 
- 2\, (x-y)\, (-1+x+y)\,\eta_i^V \, m_{l^-} \Bigg ) 
\frac{1}{L^{ver}_{h^0}}
\Bigg \}  
\, ,
\label{fVAME}  
\end{eqnarray}
where 
\begin{eqnarray}
L^{self}_{h^0}&=&m_{h^0}^2\,(1-x)+(m_i^2-m^2_{l^-}\,(1-x))\,x
\nonumber \, , \\
L^{self}_{A^0}&=&L^{self}_{1,\,h^0}(m_{h^0}\rightarrow m_{A^0})
\nonumber \, , \\
L^{ver}_{h^0}&=&m_{h^0}^2\,(1-x-y)+m_i^2\,(x+y)-q^2\,x\,y
\nonumber \, , \\
L^{ver}_{h^0\,A^0}&=&m_{h^0}^2\,x + m_i^2\,(1-x-y)+(m_{A^0}^2-q^2\,x)\,y
\nonumber \, , \\
L^{ver}_{A^0}&=&L^{ver}_{h^0}(m_{h^0}\rightarrow m_{A^0})
\nonumber \, , \\
L^{ver}_{A^0\,h^0}&=&L^{ver}_{h^0\,A^0}(m_{h^0}\rightarrow m_{A^0}) \, ,
\label{Lh0A0}
\end{eqnarray}
and 
\begin{eqnarray}
\eta_i^V&=&\xi^{D\,*}_{N,il} \xi^{D}_{N,l i}
\nonumber \, , \\
\eta_i^+&=&\xi^{D\,*}_{N,il} \xi^{D\,*}_{N,il}+
\xi^{D}_{N,li} \xi^{D}_{N,li} \nonumber \, , \\
\eta_i^-&=&\xi^{D\,*}_{N,il} \xi^{D\,*}_{N,il}-
\xi^{D}_{N,li} \xi^{D}_{N,l i}\, . 
\label{etaVA}
\end{eqnarray}
The parameters $c_V$ and $c_A$ are $c_A=-\frac{1}{4}$ and 
$c_V=\frac{1}{4}-sin^2\,\theta_W$. In eq. (\ref{etaVA}) the flavor changing
couplings $\xi^{D}_{N, l i}$ represent the effective interaction 
between the internal lepton $i$, ($i=e,\mu,\tau$) and outgoing (incoming) 
$l^-\,(l^+)$ one. It is useful to redefine the coupling as 
$\xi^{D}_{N, l i}=\sqrt{\frac{4\,G_F}{\sqrt{2}}}\, \bar{\xi}^{D}_{N, l i}$ 
to extract the dimensionfull part $\bar{\xi}^{D}_{N, l i}$.  
In general the Yukawa couplings $\bar{\xi}^{D}_{N, l i}$ are complex and 
they can be parametrized as 
\begin{eqnarray}
\bar{\xi}^{D}_{N,i l}=|\bar{\xi}^{D}_{N,i l}|\, e^{i \theta_{il}}
\,\, , 
\label{xi}
\end{eqnarray}
with lepton flavors $i,l$ and CP violating parameters $\theta_{il}$. Notice
that parameters $\theta_{il}$ are the sources of the lepton EDM. 

Finally the BR for the LFC process $Z\rightarrow l^-\, l^+$, for the vanishing
external lepton masses, can be written as 
\begin{eqnarray}
BR (Z\rightarrow l^-\,l^+)=\frac{m_Z}{12\,\pi\,\Gamma_Z}\, 
\{|f_V|^2+|f_A|^2+\frac{1}{2\,cos^2\,\theta_W} (|f_M|^2+|f_E|^2) \}\, .
\label{BR1}
\end{eqnarray}
where $f_V=f^{SM}_V+f^{Beyond}_V,\, f_A=f^{SM}_A+f^{Beyond}_A$ and 
$f_M=f^{Beyond}_M, \, f_E=f^{Beyond}_E\,$. Here $\Gamma_Z$ is the total decay 
width of Z boson, namely $\Gamma_Z=2.490 \pm 0.007\, GeV$.
\section{Discussion}
Flavor conserving $Z\rightarrow l^+ l^-$ decays are possible at the tree
level in the SM model and the contribution of one loop corrections to the 
tree level result is small. Our aim is to determine the new physics effects 
to the BR of these decays and to predict the restrictions 
for the free parameters of the model used. The model we study is the model 
III version of 2HDM, which may bring considerable contribution to the 
BR ($Z\rightarrow l^+ l^-$) beyond the SM. However, in the model III, there 
are large number of free parameters, namely, the masses of charged and neutral 
Higgs bosons, the Yukawa couplings that can be complex in general. The Yukawa 
couplings in the lepton sector are $\bar{\xi}^D_{N,ij}, i,j=e, \mu, \tau$ 
and it is necessary to restrict them using the present and forthcoming 
experiments. 

The couplings $\bar{\xi}^{D}_{N,ij},\, i,j=e,\mu $ can be neglected 
compared to $\bar{\xi}^{D}_{N,\tau\, i}\, i=e,\mu,\tau$ with the assumption
that the strength of these couplings are related with the masses of leptons 
denoted by the indices of them. Furthermore, we assume that 
$\bar{\xi}^{D}_{N,ij}$ is symmetric with respect to the indices $i$ and $j$. 
Therefore, the Yukawa couplings $\bar{\xi}^{D}_{N,\tau\, e}$, 
$\bar{\xi}^{D}_{N,\tau\, \mu}$ and $\bar{\xi}^{D}_{N,\tau\, \tau}$ play the
main role in our lepton conserving decays, $Z\rightarrow l^+ l^-$.

For $\bar{\xi}^D_{N,\mu \tau}$, the constraint coming from 
the experimental limits of $\mu$ lepton EDM \cite{Abdullah},
\begin{eqnarray}
0.3\times 10^{-19}\, e-cm < d_{\mu} < 7.1\times 10^{-19}\, e-cm 
\label{muedmex}
\end{eqnarray}
(see \cite{ErLFV} for details) or the deviation of the anomalous magnetic 
moment (AMM) of muon over its SM prediction \cite{Czarnecki} due to the 
recent experimental result of muon AMM by g-2 Collaboration \cite{Brown}, 
can be used. The coupling $\bar{\xi}^D_{N,e \tau}$ is restricted using the 
experimental upper limit of the $BR$ of the process $\mu\rightarrow e\gamma$ 
and the above constraint for $\bar{\xi}^D_{N,\mu \tau}$, since 
$\mu\rightarrow e\gamma$ decay can be used to fix the Yukawa combination 
$\bar{\xi}^{D}_{N,\mu\tau}\, \bar{\xi}^{D}_{N,e\tau}$. Using the the 
experimental bounds of $\mu$ lepton EDM and the upper limit of the $BR$ of 
the process $\mu\rightarrow e\gamma$ $|\bar{\xi}^{D}_{N,\mu\tau}|$ 
($|\bar{\xi}^{D}_{N,e\tau}|$) has been predicted at the order of the 
magnitude of $10^2-10^3$ ($10^{-5}-10^{-3}$) GeV (see \cite{ErLFV}).
For $|\bar{\xi}^{D}_{N,\tau\tau}|$ no prediction has been done yet.
 
The present work is devoted to study on the lepton flavor conserving decays 
$Z\rightarrow l^- l^+$, where $l=e,\mu,\tau$. The main contribution to the
BR of these processes come from the SM in the tree level. In the
calculations, we neglect the one loop diagrams including the charged
$W^{\pm}$ bosons in the SM and $H^{\pm}$ bosons beyond, since the neutrinos
existing in the expressions are almost massless. Therefore, we take into
account the one loop contributions including the neutral Higgs bosons $h^0$ 
and $A^0$ beyond the SM. We see that, with this additional part, 
the BR of the $Z\rightarrow l^- l^+$ decays can be enhanced, by playing 
with the Yukawa couplings and the upper limits of these Yukawa couplings 
can be restricted using the existing experimental measurements. Notice that 
in the theoretical calculations, we take the Yukawa couplings complex, 
however we use their magnitudes in the numerical analysis, since the BR of 
the processes under consideration are not sensitive to the CP violating part 
of these couplings. Throughout our calculations we use the input values given 
in Table (\ref{input}).  
\begin{table}[h]
        \begin{center}
        \begin{tabular}{|l|l|}
        \hline
        \multicolumn{1}{|c|}{Parameter} & 
                \multicolumn{1}{|c|}{Value}     \\
        \hline \hline
        $m_e     $                 & $0    $ (GeV) \\
        $m_{\mu} $                 & $0.106$ (GeV) \\
        $m_{\tau}$                 & $1.78 $ (GeV) \\           
        $m_{W}   $                 & $80.26$ (GeV) \\
        $m_{Z}   $                 & $91.19$ (GeV) \\
        $G_F     $                 & $1.16637 10^{-5} (GeV^{-2})$  \\
        $\Gamma_Z$                 & $2.490\, (GeV)$  \\
        $sin\,\theta_W$            & $\sqrt{0.2325}$ \\
        \hline
        \end{tabular}
        \end{center}
\caption{The values of the input parameters used in the numerical
          calculations.}
\label{input}
\end{table}

Now we would like to parametrize the BR as 
\begin{eqnarray}
BR=BR^{SM}+BR^{Beyond}\, ,
\label{BRpar}
\end{eqnarray}
where $BR^{SM}$ is coming from only the SM part. In eq. (\ref{BRpar}) 
$BR^{Beyond}$ gets contributions from the combination of the SM and beyond, 
which we denote as $BR^{Mixed}$, and from only beyond the SM, which we 
denote as $BR^{PureBeyond}$.

\begin{figure}[htb]
\vskip -3.0truein
\centering
\epsfxsize=6.8in
\leavevmode\epsffile{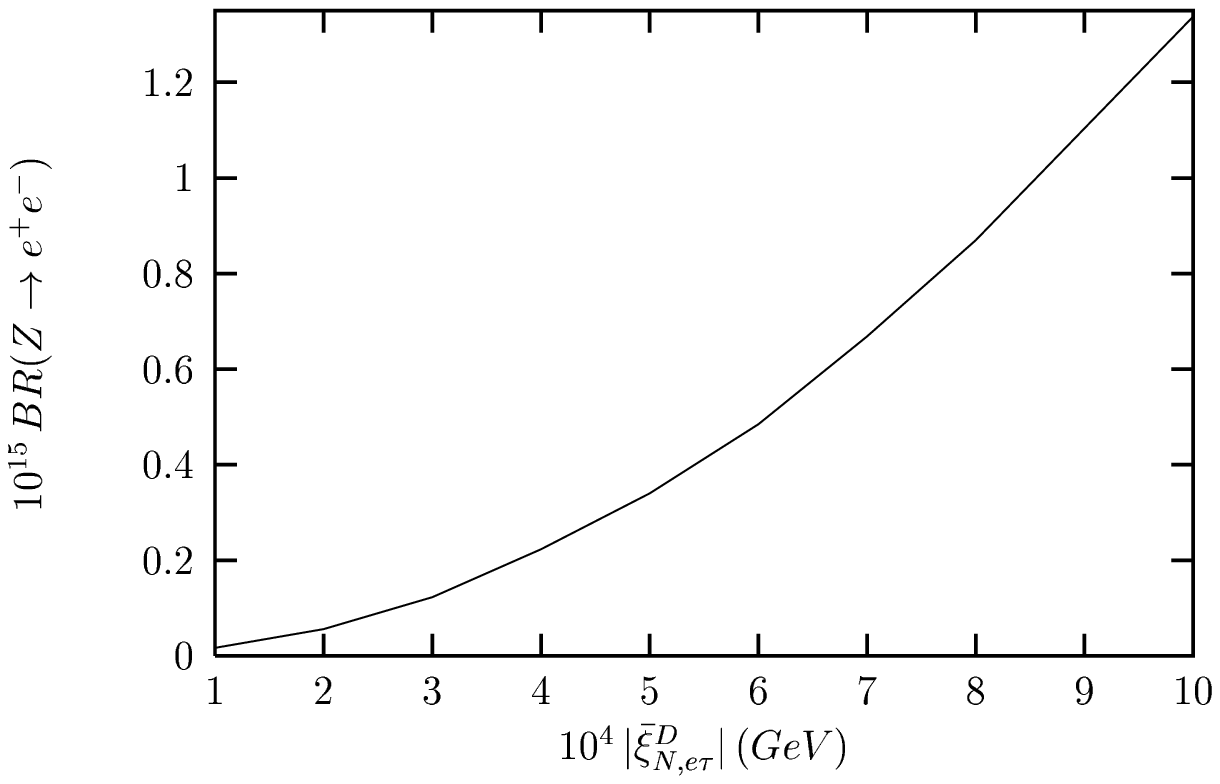}
\vskip -3.0truein
\caption[]{$|\bar{\xi}^{D}_{N,e\tau}|$ dependence of the $BR^{Beyond}\,
(Z\rightarrow e^-\, e^+)$ for $m_{h^0}=80\, GeV$ and $m_{A^0}=90\, GeV$.} 
\label{BrZeeksiDetaut}
\end{figure}

Fig. \ref{BrZeeksiDetaut} represents $|\bar{\xi}^{D}_{N,e\tau}|$
dependence of the $BR^{Beyond}\,(Z\rightarrow e^-\, e^+)$ for 
$m_{h^0}=80\, GeV$ and $m_{A^0}=90\, GeV$. Here the coupling 
$|\bar{\xi}^{D}_{N,e\tau}|$ stands in the range $10^{-4}-10^{-3}\, GeV$ 
respecting the above restrictions and the $BR^{Beyond}$ can take the 
values at most at the order of the magnitude of $10^{-15}$. This value 
is negligible compared to even the uncertainity in the experimental
result of the $BR\,(Z\rightarrow e^-\, e^+)$,  $\sim 0.008 \%$. This 
concludes that the new physics effects in the model 
III does not bring any contribution to the BR of the process 
$Z\rightarrow e^- \, e^+$ and a new constraint for the coupling 
$|\bar{\xi}^{D}_{N,e\tau}|$ can not be found. Notice that the 
$BR^{PureBeyond}\,(Z\rightarrow e^-\, e^+)$ is at the order of the magnitude 
of $10^{-28}$. We also study the Higgs boson $h^0$ mass $m_{h^0}$ dependence
of the $BR^{Beyond}$ ($50 < m_{h^0} < 80\, GeV$) and observe that, 
for $m_{h^0}=50\,GeV$, there is an enhancement more than a factor of two
larger than the result for $m_{h^0}=80\,GeV$.

\begin{figure}[htb]
\vskip -3.0truein
\centering
\epsfxsize=6.8in
\leavevmode\epsffile{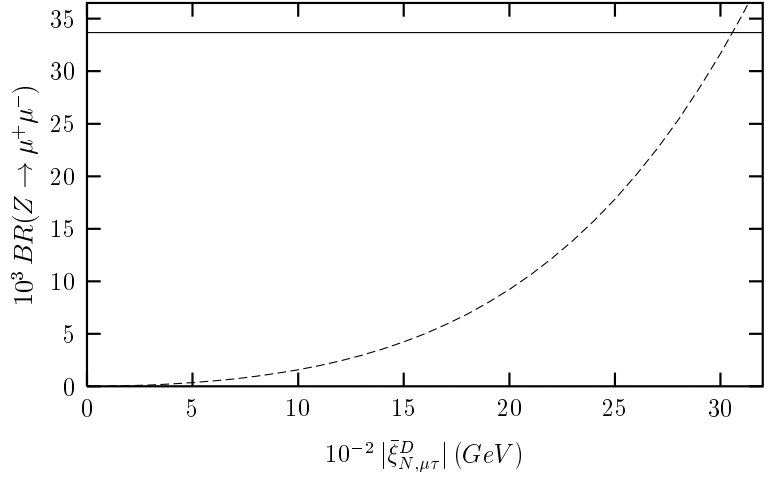}
\vskip -3.0truein
\caption[]{$|\bar{\xi}^{D}_{N,\mu\tau}|$
dependence of the $BR^{Beyond}\,(Z\rightarrow \mu^-\, \mu^+)$ for 
$m_{h^0}=80\, GeV$ and $m_{A^0}=90\, GeV$. Here the solid (dashed) line 
represents the SM (Beyond) contribution.}
\label{BrZmumuksiDmutaut}
\end{figure}

In Fig. \ref{BrZmumuksiDmutaut} we present $|\bar{\xi}^{D}_{N,\mu\tau}|$
dependence of the $BR^{Beyond}\,(Z\rightarrow \mu^-\, \mu^+)$ for 
$m_{h^0}=80\, GeV$ and $m_{A^0}=90\, GeV$. Here the solid (dashed) line 
represents the SM (Beyond) contribution. It can seen that the $BR^{Beyond}$
can reach the SM values for the large values of the coupling, 
$|\bar{\xi}^{D}_{N,\mu\tau}|\sim 3.1\times 10^3\, GeV$, which lies in the 
above constraint region. Not to exceed the experimental result of the 
process  $Z\rightarrow \mu^-\, \mu^+$, we need to predict an upper limit 
for this coupling. By taking into account the uncertainity of the 
experimental result, namely $0.013\, \%$ we get the upper limit of the 
coupling as $|\bar{\xi}^{D}_{N,\mu\tau}| \sim 3\times 10^2\, GeV$. The 
$BR^{PureBeyond}\,(Z\rightarrow e^-\, e^+)$ can also reach the SM value for
$|\bar{\xi}^{D}_{N,\mu\tau}|\sim 3.5\times 10^3\, GeV$. Furthermore, the 
Higgs boson $h^0$ mass $m_{h^0}$ dependence of the $BR^{Beyond}$ 
($50 < m_{h^0} < 80\, GeV$) shows that for $m_{h^0}=50\,GeV$, there is an 
enhancement more than a factor of three larger than the result for 
$m_{h^0}=80\,GeV$.

\begin{figure}[htb]
\vskip -3.0truein
\centering
\epsfxsize=6.8in
\leavevmode\epsffile{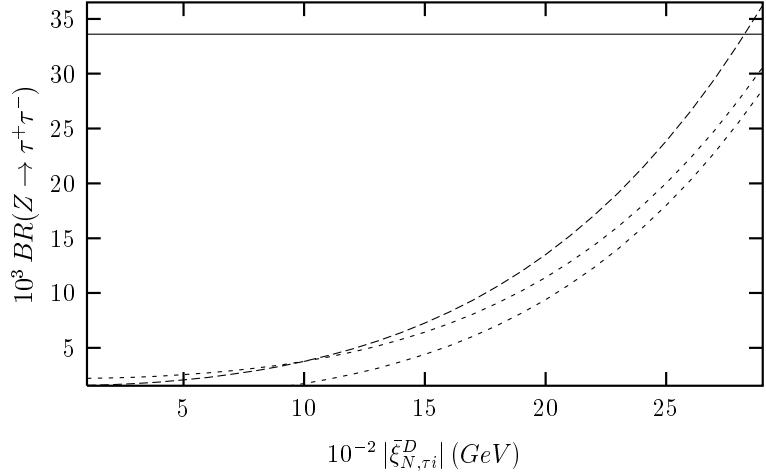}
\vskip -3.0truein
\caption[]{$|\bar{\xi}^{D}_{N,\tau i}|\, (i=\mu,\tau) $ dependence of the 
$BR^{Beyond}\,(Z\rightarrow \tau^- \, \tau^+)$ for $m_{h^0}=80\, GeV$ and 
$m_{A^0}=90\, GeV$. Here the solid line is devoted to the SM, dashed line 
to the $|\bar{\xi}^{D}_{N,\tau \mu}|$ dependence for 
$|\bar{\xi}^{D}_{N,\tau \tau}|=10^3\, GeV$, small dashed line to the 
$|\bar{\xi}^{D}_{N,\tau \tau}|$ dependence for $|\bar{\xi}^{D}_{N,\tau \mu}|
=10^3\, GeV$ and dotted line to the $|\bar{\xi}^{D}_{N,\tau \tau}|$ 
dependence for $|\bar{\xi}^{D}_{N,\tau \mu}|=3\times 10^2\, GeV$.}
\label{BrZtautauksiDtaujt}
\end{figure}

Fig. \ref{BrZtautauksiDtaujt} is devoted to the $|\bar{\xi}^{D}_{N,\tau
i}|\, (i=\mu,\tau) $ dependence of the $BR^{Beyond}\,(Z\rightarrow \tau^- 
\, \tau^+)$ for $m_{h^0}=80\, GeV$ and $m_{A^0}=90\, GeV$. Here the solid 
(dashed, small, dotted ) line represents the SM (Beyond). 
Dashed line is devoted to the $|\bar{\xi}^{D}_{N,\tau \mu}|$ dependence 
for $|\bar{\xi}^{D}_{N,\tau \tau}|=10^3\, GeV$, small dashed line is to the 
$|\bar{\xi}^{D}_{N,\tau \tau}|$ dependence for $|\bar{\xi}^{D}_{N,\tau \mu}|
=10^3\, GeV$ and dotted line is to the $|\bar{\xi}^{D}_{N,\tau \tau}|$ 
dependence for $|\bar{\xi}^{D}_{N,\tau \mu}|=3\times 10^2\, GeV$, which is
the upper limit of $|\bar{\xi}^{D}_{N,\tau \mu}|$, obtained using 
$BR^{Beyond}\,(Z\rightarrow \mu^-\, \mu^+)$.  From this figure, it is 
observed that $BR^{Beyond}$ can reach the SM value for the large values of 
the couplings, $|\bar{\xi}^{D}_{N,\mu\tau}|\sim 3.0 \times 10^3\, GeV$ and 
$|\bar{\xi}^{D}_{N,\mu\tau}|\sim 3.0 \times 10^3\, GeV$. Our aim is to get 
a $BR^{Beyond}$ so that the BR does not exceed its experimental value. To 
ensure this, first, we choose the upper limit of the coupling 
$|\bar{\xi}^{D}_{N,\mu\tau}|$  as $3\times 10^{-2}\, GeV$, respecting the 
$BR^{Beyond}\, (Z\rightarrow \mu^-\, \mu^+)$ and then use the uncertainity 
of the experimental result for $BR\, (Z\rightarrow \tau^- \, \tau^+)$, 
namely $0.015\, \%$. We predict the upper limit of the coupling as 
$|\bar{\xi}^{D}_{N,\mu\tau}| < 1\times 10^2\,GeV$ (see dotted line in the 
Fig. \ref{BrZtautauksiDtaujt}).  Notice that $BR^{PureBeyond}\,
(Z\rightarrow e^-\, e^+)$ can reach the SM value for $|\bar{\xi}^{D}_
{N,\mu\tau}|\sim 3.5\times 10^3\, GeV$ and $|\bar{\xi}^{D}_{N,\tau\tau}|
\sim 3 \times 10^3\, GeV$. For completeness we study the mass $m_{h^0}$ 
dependence of the $BR^{Beyond}$ ($50 < m_{h^0} < 80\, GeV$) and observe 
that for $m_{h^0}=50\,GeV$, there is an enhancement less than a factor of 
two larger than the result for $m_{h^0}=80\,GeV$.

As a summary, we study the $BR^{Beyond}$'s of the lepton flavor conserving 
decays $Z\rightarrow l^- l^+\, (l=e,\mu,\tau)$ in the model III. We observe
that the one loop diagrams due to neutral Higgs bosons $h^0$ and 
$A^0$ can give considerable contributions and its possible even to reach 
the tree level SM result. This forces one to predict the upper limits for 
the Yukawa couplings in the lepton sector: 
\begin{itemize}  
\item An upper limit for the coupling $|\bar{\xi}^{D}_{N,e\tau}|$ can not 
be found since the new physics effects to the $BR\,(Z\rightarrow e^- e^+)$ 
are exteremely small compared to the SM result.
\item We predict the upper limit of the coupling as 
$|\bar{\xi}^{D}_{N,\mu\tau}| < 3\times 10^2\, GeV$.
\item We predict the upper limit of the coupling as 
$|\bar{\xi}^{D}_{N,\tau\tau}| < 1\times 10^2\,GeV$.
\end{itemize}
Furthermore, $BR^{Beyond}$ is not so much sensitive to the mass $m_{h^0}$.
  
In future, with the more reliable experimental result of the $BR$'s of above 
processes it would be possible to test models beyond the SM and free 
parameters of these models
\section{Acknowledgements}
\newpage

\end{document}